\documentstyle[osa,manuscript,psfig]{revtex}
\begin{document}
\newcommand{\IP}{I\hspace{-0.04in}P}
\newcommand{\IPI}{$I\hspace{-0.04in}P$~} 

\titlepage
\title{Energy Dependence of the Contribution of Pion Exchange 
to Large-Rapidity-Gap Events in Deep Inelastic Scattering}
\author {Liu Chun-xiu and Liang Zuo-tang}
\address{Department of Physics, Shandong University, Ji'nan, Shandong
250100,China}
\maketitle

\begin{abstract}
We study the energy dependence of the
contribution of pion exchange to large-rapidity-gap events in
deep inelastic scattering. The results show that this
contribution can be quite significant at low energy 
and that the LRG events observed by E665 collaboration 
in $\mu Xe$ and $\mu D$ interactions at 490 $GeV$ 
can be reasonably well described 
in terms of meson exchange.
We also show that the distribution of the maximum rapidity
for all hadrons is quite different from 
that for charged hadrons only
and that the former exhibits also 
shoulder-like structure for events at 490 $GeV$
similar to that at HERA. 
\end{abstract}

\newpage

Large-rapidity-gap (LRG) events in deep inelastic scattering have
been studied extensively at HERA both experimentally
\cite{DHERA97,ZEUS93,H194} and 
theoretically [\ref{DHERA97},\ref{HSCI}-\ref{HJung95}]. 
It has been observed 
that there exist a class of events 
in which there are no particles or energy depositions
in the forward part of the detector,
i.e., there is a large rapidity gap in the final state.  
Such events take about $10\%$ of
the total deep inelastic events in the kinematic
range $10^{-4}<x<10^{-2}$ and $5<Q^2<120$ $GeV^2$.
(Here $x$ is the Bjorken-$x$,
$Q^2=-q^2$ and $q$ is the four-momentum transfer 
carried by the virtual photon.)
The experimental data
obtained by ZEUS \cite{ZEUS93} and H1 \cite{H194} collaborations
show that the distribution of
the maximum pseudo-rapidity $\eta_{max}$ 
for the produced hadrons 
has a clear shoulder-like structure
which signifies the occurrence of 
the LRG events.

The existence of the LRG shows that, 
in such events, the exchanged object
between the virtual photon and 
the incoming proton must be colorless 
(c.f. Fig. 1a given later in this paper).
It has been shown that different features of LRG events 
can be well described \cite{H194} 
in terms of pomeron (\IPI) exchange 
in Regge phenomenology. 
At the same time, it was also well known that 
one-meson exchange such as 
one-pion exchange process 
contributes significantly \cite{SULL72} to
deep inelastic scattering $ep\to eX$.
One-pion exchange itself 
takes about $10\%$ of the whole $ep\to eX$ events
in the abovementioned kinematic region \cite{BL95}.
Therefore, it was expected \cite{BL95} that 
meson exchange might also contribute to the LRG events.
This can be studied using Monte Carlo event generators.
Such study showed that the contribution
is negligible \cite{pionTH} at HERA energy. 
It has completely no contribution to 
the characteristic shoulder-like structure
in $\eta_{max}$ distribution.

It is interesting to note that
LRG events have also been observed \cite{E66595} 
by E665 collaboration at FNAL
in $\mu Xe$ and $\mu D$ fixed target experiments at 490 $GeV$. 
In these experiments, 
LRG events were defined as the events
with a rapidity gap $\Delta y^*>2$,  
where $\Delta y^*$ is
the difference between the rapidity of the
target nucleon before the scattering  
and the lowest rapidity of the
charged hadrons in the event in 
$\gamma^*$-nucleon c.m. frame.  
The distribution for the probability of $\Delta y^*$
greater than a given value 
i.e., the probability distribution $P(\Delta y^*)$,
rather than the probability density,
has been given.
It is interesting to note 
that the $P(\Delta y^*)$ 
obtained by E665 does not 
have the same shoulder-like
structure as that for $\eta_{max}$ distribution 
at HERA\cite{ZEUS93,H194}. 
Having in mind that $\Delta y^*$ and $\eta_{max}$
are essentially the same 
except for the small difference in reference system,
we are naturally led to the following questions:
Why is there such a large difference between 
the $\Delta y^*$ distribution obtained by E665 collaboration
and the $\eta_{max}$ distribution 
obtained by H1 and ZEUS collaborations?
How is the energy dependence of the rapidity distribution in events 
where \IPI  or $\pi$ is exchanged?
Can meson exchange give a significant contribution 
to the LRG events observed by E665 collaboration?

These are the questions that 
we would like to study in this note.
For explicity, 
we take one-pion exchange as an example. 
Similar effects should also exist for other mesons. 
We now start with a qualitative analysis.
In fact, from the following simple qualitative analysis, 
we already expect that 
the contribution of 
\IPI or $\pi$ exchange to LRG events at E665 energy 
may be quite different from that at HERA energy.
We use the notations as those shown in Fig.1a
and recall that the invariant mass of 
the total hadronic system and that 
of the hadronic subsystem $X$ are given by,
\begin{equation}
W^2\equiv(P+q)^2=Q^2(1/x-1)+M^2,
\label{eq:W2}
\end{equation}
\begin{equation}
M_X^2\equiv(q+q_{\kappa})^2=Q^2(\xi/x-1)+t,
\label{eq:MX2}
\end{equation}
respectively. Here $t\equiv q_\kappa^2$ is the square of the
four-momentum transfer between the proton and the virtual photon; 
$\xi\equiv (q\cdot q_{\kappa})/(q\cdot P)$ 
can be interpreted as the fraction of the momentum carried 
by the exchanged object $\kappa$ 
(which represents \IPI or $\pi$) 
from the incident proton
in the infinite momentum frame. 
Independent of the reference frame one uses,
the width of the rapidity distribution 
of the hadronic subsystem $X$ is proportional to $\ln M^2_X$,
i.e., $\Delta y_X\sim \ln M^2_X$;
and the width of the rapidity distribution 
of the total hadronic system 
is proportional to $\ln W^2$.
This is illustrated in Fig.1b.
We see that,  for a given $W$, the smaller $M_X$ is,
the larger the rapidity gap between the hadronic subsystem $X$
and the scattered nucleon is.
Hence, $M_X \ll W$ is a necessary condition 
for LRG to appear.
From Eq. (\ref{eq:MX2}), we see that 
$M_X$ increases with increasing $\xi$ 
and decreases with increasing $|t|$ 
at fixed $Q^2$ and $x$.
This implies that $\xi$ can not be large to ensure $M_X\ll W$.
In the experiments at HERA\cite{ZEUS93,H194},
the kinematic range is $5<Q^2<120$ $GeV^2$ and $0.0001<x<0.01$,
and the typical value for $M_X$ is several ($1\sim 7$, say) $GeV$. 
The corresponding $\xi$ value is of
the order of 0.01 or less.
This means that only events where $\xi$ is very small 
contribute to the LRG events at HERA.
Hence, the contribution of $\pi$ or \IPI exchange 
to LRG events at HERA
is determined by its contribution to deep inelastic scattering
at small $\xi$.

The $\xi$-dependences of 
the contributions of \IPI and $\pi$ exchange 
to deep inelastic scattering $ep \to eX$ 
at given $x$ and $Q^2$
are given by their contributions 
to the ``diffractive structure function'' $F_2^{D(4)}(x,Q^2,\xi,t)$. 
To make a qualitative comparison of 
the two contributions, 
we use the approximately valid factorization theorem, 
i.e. take  
$F_{2(\kappa)}^{D(4)}(x,Q^2,\xi,t)=$
$f_{\kappa/p}(\xi,t)F_2^\kappa(x/\xi,Q^2)$. 
Here, $f_{\kappa/p}$ 
is the corresponding flux factor; 
$F_2^\kappa$ 
is the structure function of $\kappa$ (\IPI or $\pi$).
In the case of \IPI exchange, 
$f_{\IP/p}(\xi,t)$ can be parameterized as\cite{DL87}
\begin{equation}
f_{\IP/p}(\xi, t)=\frac{9\delta^2}{4\pi^2}[F(t)]^2
\xi^{1-2\alpha _{\IP} (t)},
\label{eq:IPflux}
\end{equation}
where $\delta^2=3.24$ $GeV^{-2}$,
$\alpha _{\IP}(t)=1+\epsilon+\alpha't$,
$\epsilon\approx 0.085$, $\alpha'=0.25$
and the elastic form factor $F(t)$ is given by:
\begin{equation}
F(t)=\frac{4m^2_p-2.8t}{4m^2_p-t}\frac{1}{(1-t/0.7)^2},
\label{eq:ft}
\end{equation}
where $m_p$ is the mass of proton. 
The pion flux factor can be derived 
from the pion cloud model.
For the case $p\rightarrow \pi^* N$, 
we have\cite{pionTH},
\begin{equation}
f_{\pi/p}(\xi,t)=3.257\xi \frac{-t}{(m^2_{\pi}-t)^2}
\exp\left(-\frac{m^2_{\pi}-t}{1.21\xi}\right),
\label{eq:piflux}
\end{equation}
where both $m_\pi^2$ and $t$ are taken in unit of GeV$^2$. 
Take,
\begin{equation}
F_2^{\IP}(x,Q^2)=3x(1-x)/2,
\end{equation}
and the SMRS-P2-parameterization of $F_2^\pi(x,Q^2)$, 
we obtained their contributions to 
$F_2^{D(4)}(x,Q^2,\xi,t)$ 
at $t=-|t|_{min}\approx$ $m_p^2\xi^2/(1-\xi)$ in Fig.2.    
From the figure, we explicitly see that
$F_{2(\IP)}^{D(4)}(x,Q^2,\xi,-|t|_{min}) \gg 
 F_{2(\pi)}^{D(4)}(x,Q^2,\xi,-|t|_{min})$
in the small $\xi$ (say, $\xi < 0.05$) region.
Since the LRG events at HERA come mainly 
from this small $\xi$ region,
we reach the conclusion that
the contribution of \IPI exchange to LRG events at HERA
is dominant and that the contribution of $\pi$ exchange 
can be neglected.

The situation is, however, quite different
in the E665 experiment.  
Here, events in the kinematic range
$1<Q^2<100$ $GeV^2$ 
and $0.002<x<0.3$ were selected\cite{E66595}. 
Compared with those at HERA,
$Q^2$ is much smaller and $x$ is larger.  
Thus, to obtain the same $M_X$ as that at HERA,
$\xi$ should be much larger. 
In fact, for typical $M_X$ around $1\sim 7$ $GeV$,
$\xi$ is of the order of $10^{-1}\sim 10^{-2}$
and can even be significantly larger than 0.1.
From Fig.2,
we see that $F_{2(\IP)}^{D(4)}(x,Q^2,\xi,-|t|_{min})$
decreases with increasing $\xi$,
but $F_{2(\pi)}^{D(4)}(x,Q^2,\xi,-|t|_{min})$  
increases very rapidly with increasing $\xi$.
As a result, their difference becomes very small 
in the region of $\xi\sim 0.05$,
and the latter 
can even be larger than the former 
for large $\xi$ ($>0.1$, say).
Thus the contribution of pion exchange to the LRG events 
obtained in the E665 experiments 
should be quite significant 
compared with that from Pomeron.

Now, we explicitly calculate 
the rapidity distribution of final hadrons in events
where \IPI or $\pi$ exchange takes place. 
Presently this can only be carried out using Monte Carlo events generator. 
There exist a number of Monte Carlo event generators,
such as P{\scriptsize OMPYT}\cite{Bruni96} 
and R{\scriptsize APGAP}\cite{HJung95},
which simulate the processes shown in Fig.1a. 
Here, it is envisaged that the incoming proton 
`emits' a \IPI or a $\pi^*$,
and the \IPI or $\pi^*$ then 
collides with the virtual photon 
emitted by incident lepton
to produce the sub-hadronic-system $X$
shown in Fig.1a. 
Various options for the effective \IPI flux
and the parton densities 
in \IPI are available in the programs.
Using such Monte Carlo program
we can easily calculate the contribution of \IPI or $\pi$ exchange
to the LRG events at different energies 
or in different reference frames.
Both P{\scriptsize OMPYT} and R{\scriptsize APGAP} 
are slave systems, which must be called by our own
steering program.
Thus we first simulate the events at HERA to check 
our steering program then 
apply it to E665 energy.

We simulate the events where the abovementioned 
\IPI or $\pi$ exchange takes place using P{\scriptsize OMPYT} 
or R{\scriptsize APGAP}
and the usual DIS events using L{\scriptsize EPTO}\cite{Lep61}
and obtain the $\eta_{max}$ distribution 
for each class of events respectively.
We calculated them using different options for 
\IPI structure functions. 
We found out that 
both the results from P{\scriptsize OMPYT} 
and those from R{\scriptsize APGAP} 
with different options for \IPI structure function
are essentially the same.  
Adding the different contributions 
together with the corresponding weights
which measure the relative contribution of 
each type of process to the inclusive process $ep\to eX$,
we can obtain the $\eta_{max}$ distribution 
for final hadrons in deep inelastic scattering at HERA energy.  
The contribution of $\pi$ exchange 
to deep inelastic scattering $ep\to eX$
in the HERA kinematic region was estimated in [\ref{BL95}].
The results depend on the parameterization 
of pion structure function, but they are all 
of the order of $10\%$ of all DIS events at HERA energy.
In Fig.3a, we show the results that we obtained 
by adding 12\% from \IPI or $\pi$ exchange with 
88\% usual DIS events from L{\scriptsize EPTO}  
(the lower solid and dotted lines). 
Here, in obtaining these results,
all hadrons that can be observed by the H1 detector,
i.e., those with $-3.8<\eta<3.65$ 
and energy higher than 400 $MeV$,
are taken into account.
Since our purpose is to study the 
energy dependence of the contribution
of \IPI or $\pi$ exchange to LRG events, 
we do not simulate the detector effects at HERA.
The results show that 
the usual DIS events can give a reasonable account of 
the shape of the $\eta_{max}$ distribution for values above 1.5
and that the usual DIS and \IPI exchange together can well
describe the $\eta_{max}$ distribution at HERA\cite{H194}.
The results also show that the $\pi$
exchange has a very small contribution to the LRG events at HERA.

Subsequently, we apply the method 
to the $\mu Xe$ and $\mu D$ fixed target scattering 
at 490 $GeV$ beam energy\cite{E66595},
and calculate the rapidity distribution of final hadrons in events 
where \IPI or $\pi$ exchange takes place.
Since our purpose is to study 
the energy dependence of the contribution of 
\IPI or $\pi$ exchange to the LRG events,
we do not take the nuclear effects into account.
This means that we simply regard 
$\mu D$ or $\mu Xe$ scattering in the E665 kinematic region
as $\mu N$ scattering in the same kinematic region.
We select the events in the same kinematic range 
as that chosen by E665 collaboration \cite{E66595}
and obtain the probability distributions of $\Delta y^*$ 
in events where \IPI (upper solid line) 
or $\pi$ (upper dotted line) exchange takes place
compared with the E665 $\mu D$ or $\mu Xe$ data\cite{E66595}
shown in Fig.4a.
From the results we see clearly that,
as we expected in the abovementioned qualitative analysis, 
$\pi$ exchange can have a very significant contribution \cite{Trip} 
to the LRG events observed by E665 collaboration. 
We found out also, to reproduce the E665 data\cite{E66595}, 
we need a rather large contribution 
of \IPI and/or $\pi$ exchange.
We estimated the contribution of 
pion-exchange to $\mu p\to\mu X$ 
using the pion flux factor given in Eq.(\ref{eq:piflux}) 
and different parameterizations of pion 
structure functions. 
We found out that the results are slightly different 
if different parameterizations are used.  
But they are all of the order of $10\sim20\%$
in the E665 kinematic regions. 
In Fig.4a, we show the results 
obtained by adding 20\% \IPI exchange 
with 80\% from L{\scriptsize EPTO} (lower solid line)
and those obtained by adding 20\% $\pi$ exchange 
with 80\% from L{\scriptsize EPTO} (lower dotted line).
We see in particular that pion exchange 
contributes significantly to the LRG events 
but cannot account for all of them. 
However, what we discussed till now is only 
the contribution from the case $p\to\pi^* N$, 
which is an explicit and calculable 
example of different meson exchange processes.
Similar contributions should be expected 
from other mesons which cannot be calculated 
presently because of 
the lack of the corresponding flux factors 
and the structure functions.  
To show what we may expect 
if all different meson exchange processes 
are taken into account, 
we simply add more contributions from 
$\pi$ exchange to the whole events. 
Hence, in Fig.4a, we show also the results 
obtained by adding 40\% from $\pi$ exchange with 
60\% from L{\scriptsize EPTO} (dash-dotted line).
We see that the results 
agree reasonably well with the data.

From Fig.4a, we also see that the shape 
of the $\Delta y^*$ distribution does not have 
a clear shoulder-like structure at the E665 energy
as that in $\eta_{max}$ distribution 
at HERA (see Fig.3a).
We are therefore led to the question about  
the reason of the disappearance of such
characteristic structure for LRG events.
We note that,
besides the energy is lower,
only the charged hadrons are taken into account by E665
whereas all the hadrons are taken into account at HERA. 
Thus we take also all the hadrons 
with energy higher than 400 $MeV$ into account 
and re-calculate $\Delta y^*$ distribution at E665 energy.
The obtained results are shown in Fig.4b. 
It is interesting to see that,
compared with that for charged hadrons only,
the $\Delta y^*$ distribution obtained from L{\scriptsize EPTO} 
is much narrower
whereas those obtained from R{\scriptsize APGAP}
for $\pi$ or \IPI exchange remain essentially the same.
Their difference becomes much larger.
Adding them together with the corresponding weights 
mentioned above,
we obtain the total $\Delta y^*$ distribution, 
which has now a significant shoulder-like structure 
in the case of \IPI exchange. 
But the shoulder structure is not clear if only 
$\pi$ exchange is involved. 
This is similar to that at HERA.
These results clearly show that it is
much more efficient to distinguish events 
where \IPI or $\pi$ is exchanged 
from the usual DIS events 
by studying the $\Delta y^*$ distribution
for all the hadrons than that for charged hadrons only.
To check whether this is also true at other energies,
we re-calculate the $\eta_{max}$ distribution 
for the charged hadrons only at HERA energy.
The obtained result is shown in Fig.3b.
The results show similar effect, i.e.,
compared with those for all the hadrons (see Fig.3a),
the $\eta_{max}$ distribution
for the charged hadrons from usual DIS or $\pi$ exchange events 
at HERA is much wider
whereas that for \IPI exchange remains essentially the same.
The contamination from the fluctuation in usual DIS
to LRG events would be much higher 
if one would study charged hadrons only.

We note also that one can use different variables to 
describe LRG events,
such as the $\Delta y^*$ used by E665,
$\eta_{max}$ used at HERA
or $\eta^*_{max}$ and $y^*_{max}$ in the hadronic c.m. system.
We also studied the question of 
which one is more efficient by 
calculating the corresponding distributions
in the scattering of the 490 $GeV$ lepton beam  
off the fixed target proton.
The obtained results show  
no significant difference between these variables,
all of them can give good description to
the occurrence of LRG events.

In summary, using the Monte Carlo event 
generators P{\scriptsize OMPYT}, R{\scriptsize APGAP}
and L{\scriptsize EPTO}, 
we showed that
$\pi$ exchange has a significant contribution to the LRG
events in the $\mu Xe$ and $\mu D$ fixed target scattering
at 490 $GeV$ beam energy.  
Taking all the contributions from 
different meson exchange processes into account, 
we should obtain a reasonably well description of 
the corresponding E665 data at that energy.  
The distribution of the maximum rapidity
for all hadrons is quite different from 
that for charged hadrons only
and the former exhibits also 
shoulder-like structure for events at 490 $GeV$ beam energy
similar to that at HERA. 

\vskip 0.2cm

We are in debated to C. Boros who took part in 
the early stage of this research. 
We thank Li Shi-yuan, Xie Qu-bing, 
and Xu Qing-hua for discussions.  
This work was supported in part by the National Science Foundation 
(NSFC) and the Education Ministry of China.

\newpage

\begin{figure}[ht]
\psfig{file=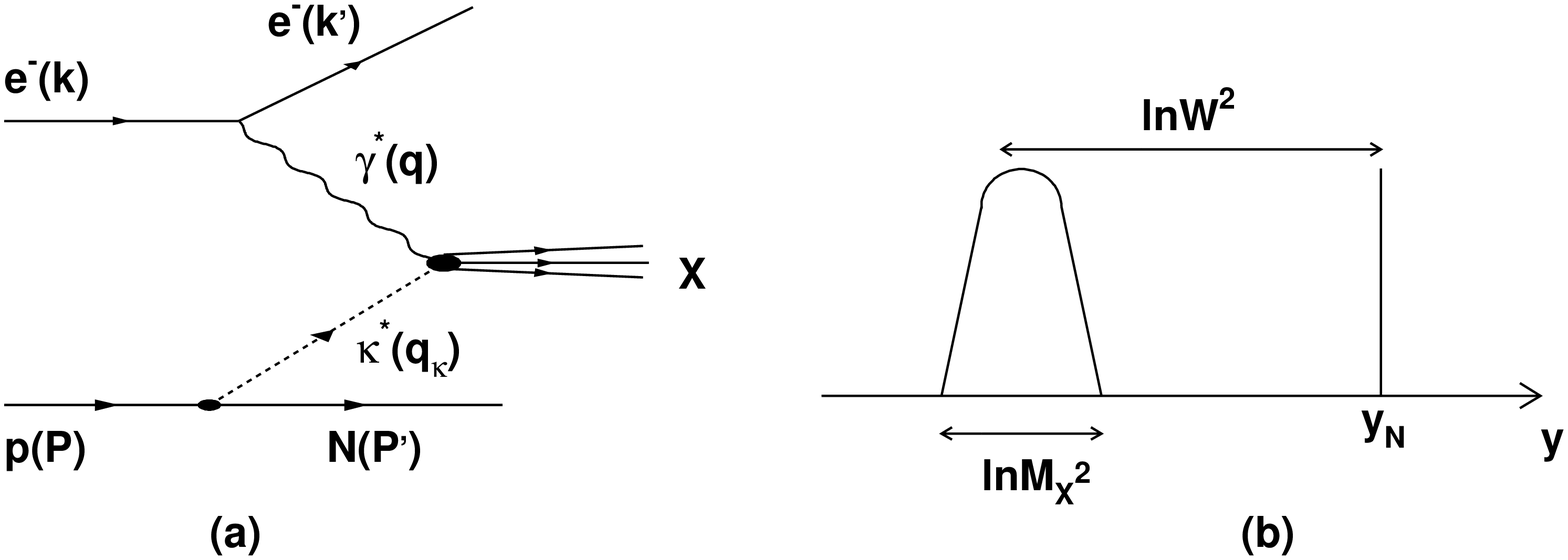,width=10cm}
\caption{(a) Schematic of the semi-inclusive process $ep\to eNX$
with the exchange of a colorless object $\kappa^*$
which can be a \IPI or a $\pi^*$; and 
(b) Diagram illustrating the range of the rapidities of
the hadrons in the subsystem $X$
and that for all the hadrons including the outgoing nucleon $N$.
Here we see in particular that the necessary condition
for LRG to appear is $\ln W^2\gg \ln M^2_X$.}
\label{fig1}
\end{figure}

\begin{figure}[ht]
\psfig{file=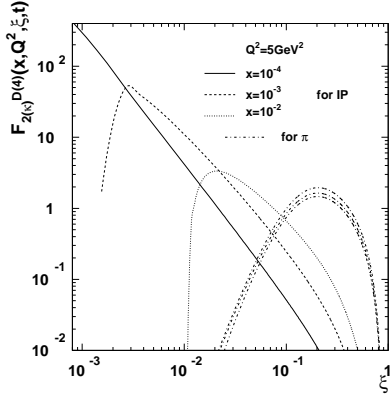,width=6cm}
\caption{$\xi$-dependence of
\IPI or $\pi$ exchange contribution to 
the ``diffractive structure function'' 
$F_{2(\kappa)}^{D(4)}(x,Q^2,\xi,t)$ at $t=-|t|_{min}$.}
\label{fig2}
\end{figure}

\newpage 

\begin{figure}[ht]
\psfig{file=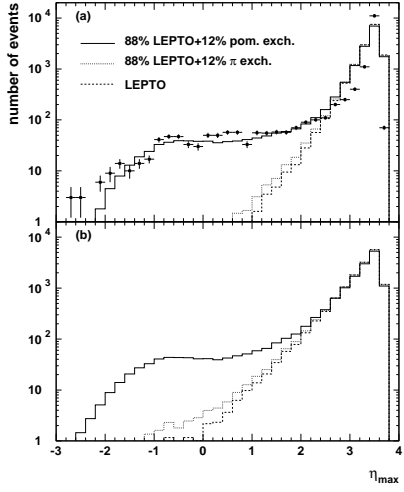,width=6cm}
\caption{Distribution of $\eta_{max}$
for hadrons in DIS events at HERA. 
The data were taken from [\ref{H194}].
The total number of the Monte Carlo
events is normalized to the data.
In obtaining the results in (a), 
all hadrons which can be observed
by H1 detector (i.e. those with $-3.8<\eta<3.65$ 
and energy higher than 400 MeV) 
are taken into account.
In (b), only the charged hadrons
are taken into account.}
\label{fig3}
\end{figure}

\newpage 

\begin{figure}[ht]
\psfig{file=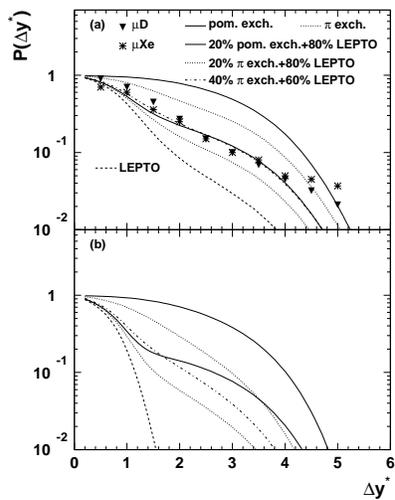,width=6cm}
\caption{Probability $P(\Delta y^*)$ 
of events with rapidity difference between 
the most backward hadron and the incoming nucleon 
in the $\gamma^*p$ c.m. frame to be greater 
than $\Delta y^*$ in the scattering of 
490 $GeV$ lepton off fixed proton target.
In obtaining the results in (a), 
only the charged hadrons with energy 
higher than 400 $MeV$ are taken into account.
The data are obtained by E665 collaboration
[\ref{E66595}] in $\mu Xe$ and $\mu D$ 
scattering at 490 GeV.
In (b), all the hadrons 
with energy higher than 400 $MeV$ are taken into account.}
\label{fig4}
\end{figure}

\newpage 

Figure captions

\vskip 0.3cm

\noindent 
Fig.1: (a) Schematic of the semi-inclusive process $ep\to eNX$
with the exchange of a colorless object $\kappa^*$
which can be a \IPI or a $\pi^*$; and 
(b) Diagram illustrating the range of the rapidities of
the hadrons in the subsystem $X$
and that for all the hadrons including the outgoing nucleon $N$.
Here we see in particular that the necessary condition
for LRG to appear is $\ln W^2\gg \ln M^2_X$.

\vskip 0.3cm
\noindent 
Fig.2: $\xi$-dependence of
\IPI or $\pi$ exchange contribution to 
the ``diffractive structure function'' 
$F_{2(\kappa)}^{D(4)}(x,Q^2,\xi,t)$ at $t=-|t|_{min}$.

\vskip 0.3cm
\noindent 
Fig.3: Distribution of $\eta_{max}$
for hadrons in DIS events at HERA. 
The data were taken from [\ref{H194}].
The total number of the Monte Carlo
events is normalized to the data.
In obtaining the results in (a), 
all hadrons which can be observed
by H1 detector (i.e. those with $-3.8<\eta<3.65$ 
and energy higher than 400 MeV) 
are taken into account.
In (b), only the charged hadrons
are taken into account.

\vskip 0.3cm
\noindent 
Fig.4: Probability $P(\Delta y^*)$ 
of events with rapidity difference between 
the most backward hadron and the incoming nucleon 
in the $\gamma^*p$ c.m. frame to be greater 
than $\Delta y^*$ in the scattering of 
490 $GeV$ lepton off fixed proton target.
In obtaining the results in (a), 
only the charged hadrons with energy 
higher than 400 $MeV$ are taken into account.
The data are obtained by E665 collaboration
[\ref{E66595}] in $\mu Xe$ and $\mu D$ 
scattering at 490 GeV.
In (b), all the hadrons 
with energy higher than 400 $MeV$ are taken into account.


\vskip 0.2cm
 
\noindent
\begin{thebibliography}{2000}
\baselineskip 12pt
\bibitem{DHERA97} There exist already a number of reviews
  on this topic, e.g., Nicol\'{o} Cartiglia, 
  Lecture given at the SLAC summer school, 1996, 
  hep-ph/9703245 (1997). 
\label{DHERA97}
\bibitem{ZEUS93} M. Derrick {\it et al}. (ZEUS Coll.), Phys.
        Lett. B{\bf 315}, 481 (1993); 
        Z. Phys. C{\bf 68}, 569 (1995).
\label{ZEUS93}
\bibitem{H194} T. Ahmed {\it et al}. (H1 Coll.), Nucl. Phys.
        {\bf B429}, 477 (1994); Phys. Lett {\bf 348}, 681 (1995).
\label{H194}
\bibitem{HSCI} G. Ingelman, P.E. Schlein, Phys. Lett. B{\bf 152}, 256 (1985);
G. Ingelman, K. Prytz, Z. Phys. C{\bf 58}, 285(1993);
A. Edin, G. Ingelman and J. Rathsman, Phys. Lett. B{\bf 366}, 371 (1996);
Z. Phys. C{\bf 75}, 57 (1997).
\label{HSCI}
\bibitem{Niko92} N. Nikolaev, B. Zakharov, Z. Phys. C{\bf 53}, 331 (1992); 
J.~Bartels, H. Lotter, M. W\"usthoff, Phys. Lett. B{\bf 379}, 239 (1996); 
W. Buchmuller, M. McDermott, A. Hebecker, Nucl. Phys. B{\bf 487}, 283 (1997); 
M. W\"usthoff, Phys. Rev. D{\bf 56}, 4311 (1997).
\label{Niko92}
\bibitem{Bruni96} P. Bruni and G. Ingelman, Proc. of the
        Inter. Europhys. Conf. on High Energy
        Phys., eds. J. Carr and M. Perrotet, Marseille, 1993 
       (Ed. Frontieres, Gif-sur-Yvette, 1994), p.595;
	and the references given therein.
\label{Bruni96}
\bibitem{HJung95} H. Jung, Comp. Phys. Commun. {\bf 86}, 147 (1995).
\label{HJung95}
\bibitem{SULL72} J. D. Sullivan, Phys. Rev. {\bf D5}, 1732 (1972).
\label{SULL72}
\bibitem{BL95} C. Boros and Z. Liang, 
                Phys. Rev. D{\bf 51}, R4615 (1995).
\label{BL95}
\bibitem{pionTH} H. Holtmann {\it et al}., Phys. Lett.
		B{\bf 338}, 363 (1994).
\label{pionTH}
\bibitem{E66595} M. Adams {\it et al}. (E665 Coll.),
        Z. Phys. C{\bf 65},225(1995).
\label{E66595}
\bibitem{DL87} A. Donnachie, P.V. Landshoff, 
              Phys. Lett. B{\bf 191}, 309 (1987); 
              Nucl. Phys. B{\bf 303}, 634 (1988).
\label{DL87}
\bibitem{Lep61} G. Ingelman, Proc. ``Physics at
        HERA'', Eds. W. Buchmueller {\it et al.}, DESY Hamburg 1992,
        vol.3 p.1366.
\label{Lep61}
\bibitem{Trip} It should be interesting to mention that,  
 the $\pi$ flux obtained 
 in triple-Regge formalism (See [\ref{FF74}])
 dies out not so strongly at $\xi\to0$  
 compared with that given in Eq.(\ref{eq:piflux}) 
 which we used in obtaining the results in Fig.4. 
 It can be expected that the resulting $\Delta y^*$ 
 distribution should be even closer to \IPI exchange 
 if such $\pi$ flux is used. 
 To see how large the difference is, 
 we did the calculations 
 using R{\scriptsize APGAP} 
 by replacing the $\pi NN$ form factor with  
 the corresponding results in [\ref{FF74}].    
 We found that the obtained $\Delta y^*$ distribution 
 is indeed closer to that from \IPI exchange 
 but the differences 
 between the results using these two different 
 pion fluxs are not very large. 
\label{Trip} 
\bibitem{FF74} R.D. Field and G.C. Fox, 
    Nucl. Phys. B{\bf 80}, 367 (1974).
\label{FF74}
\end{thebibliography}
\end{document}